# Maximising and Stabilising Luminescence in Metal Halide Perovskite Device Structures

Mojtaba Abdi-Jalebi[1], Zahra Andaji-Garmaroudi[1], Stefania Cacovich[2], Camille Stavrakas[1], Bertrand Philippe[3], Johannes M. Richter[1], Mejd Alsari[1], Edward P. Booker[1], Eline M. Hutter[4], Andrew J. Pearson[1], Samuele Lilliu[5], Tom J Savenije[4], Håkan Rensmo[3], Giorgio Divitini[2], Caterina Ducati[2], Richard H. Friend[1], Samuel D. Stranks[1]*

[1]Cavendish Laboratory, Department of Physics, University of Cambridge, JJ Thomson Avenue, Cambridge CB3 0HE, UK

[2]Department of Materials Science & Metallurgy, University of Cambridge, 27 Charles Babbage Road, Cambridge CB3 0FS, UK

[3]Department of Physics and Astronomy, Uppsala University, Box 516, 75120 Uppsala, Sweden

[4]Opto-electronic Materials Section, Department of Chemical Engineering, Delft University of Technology, van der Maasweg 9, 2629 HZ Delft, The Netherlands

[5]Department of Physics and Astronomy, University of Sheffield, Sheffield, S3 7RH, UK

*sds65@cam.ac.uk

**Metal halide perovskites are attracting tremendous interest for a variety of high-performance optoelectronic applications[1]. The ability to continuously tune the perovskite bandgap by tweaking the chemical compositions opens up new applications for perovskites as coloured emitters, in building-integrated photovoltaics, and as components of tandem photovoltaics to further increase the power conversion efficiency[2–4]. Nevertheless, parasitic non-radiative losses are still limiting performance, with luminescence yields in state-of-the-art perovskite solar cells still far from 100% under standard solar illumination conditions[5–7]. Furthermore, in mixed halide perovskite systems designed for continuous bandgap tunability (bandgaps ~1.7-1.9 eV)[2], photo-induced ion segregation leads to bandgap instabilities[8,9]. Here, we substantially mitigate both non-radiative losses and photo-induced ion migration in perovskite films and interfaces by decorating the surfaces and grain boundaries with passivating potassium-halide interlayers. We demonstrate external photo-luminescence quantum yields of 66%, translating to internal yields exceeding 95%. The high luminescence yields are achieved while maintaining high mobilities over 40 $cm^2V^{-1}s^{-1}$, giving the elusive combination of both high luminescence and excellent charge transport[10]. We find that the external luminescence yield when interfaced with electrodes in a solar cell device stack, a quantity that must be maximized to approach the efficiency limits, remains as high as 15%, indicating very clean interfaces. We also demonstrate the inhibition of transient photo-induced ion migration processes across a wide range of mixed halide perovskite bandgaps that otherwise show bandgap instabilities. We validate these results in full operating solar cells, highlighting the importance of maximising and stabilising luminescence in device structures. Our work represents a critical breakthrough in the construction of tunable metal halide perovskite films and interfaces that can approach the efficiency limits in both tandem solar cells, coloured LEDs and other optoelectronic applications.**

We fabricate a series of passivated triple-cation perovskite thin films on glass[11] (Cs$_{0.06}$FA$_{0.79}$MA$_{0.15}$Pb(I$_{0.85}$Br$_{0.15}$)$_3$, where MA = methylammonium, CH$_3$NH$_3$; FA = formamidinium, CH$_3$(NH$_2$)$_2$, by diluting the precursor solution with KI solution. We herein denote the perovskite as (Cs,FA,MA)Pb(I$_{0.85}$Br$_{0.15}$)$_3$ and the passivated samples with $x$ = [K]/([A]+[K]) and A = (Cs,FA,MA), where $x$ represents the fraction of K out of the A-site cations in the precursor solution. We note that the standard triple-cation precursor solution recipe ($x = 0$) has a slight halide deficiency but introducing KI leads to samples with a small excess of halide, along with very slight changes to the I/Br ratio (Extended Data Figure 1). The films have uniformly packed grains each of size of ~200-400 nm (Extended Data Figure 2). Absorption and photoluminescence measurements reveal a reduction in the optical bandgap of the perovskite film with increasing KI addition, consistent with the additives selectively interacting with the bromide (Extended Data Figures 3 and 4).

For a solar cell or light-emitting diode to approach its efficiency limits, all recombination should be radiative and the luminescence efficiency should be maximised[12]. In state-of-the-art perovskite films, there are still substantial non-radiative losses originating from sub-gap charge-carrier trap states present in the perovskite layer[13], which have been reported to be particularly numerous on grain surfaces[14]. The origin of the sub-gap states is still unclear, but they may be associated with ionic defects such as halide vacancies[15,16]. In Figure 1a, we show the external PL quantum efficiency (PLQE) of the (Cs,FA,MA)Pb(I$_{0.85}$Br$_{0.15}$)$_3$ perovskite films with increasing K content measured at excitation densities equivalent to solar illumination conditions. The PLQE shows a significant jump from the initial value of 8% ($x = 0$) to 41% ($x = 0.05$) reaching a remarkably high PLQE of 66% for $x = 0.40$. By accounting for photon recycling and poor light-out-coupling effects[17], these values translate to an internal PLQE exceeding 95% for the passivated compositions (Figure 1a). Furthermore, an intensity-dependence for the samples reveals that the PLQE does not change significantly with excitation power for the K-samples, unlike the $x = 0$ sample in which the PLQE increases with intensity due to the presence of high trap densities that can be filled with charge carriers at higher fluences[18] (Extended Data Figure 3). These results are also reflected in micro-PL measurements (Extended Data Figure 5). Time-resolved PL (TRPL) measurements (Figure 1b) show an enhancement in the carrier lifetime for samples with the K-based additives due to the elimination of the fast non-radiative component seen in $x = 0$, which leads to radiative bimolecular recombination (Extended Data Figure 6).

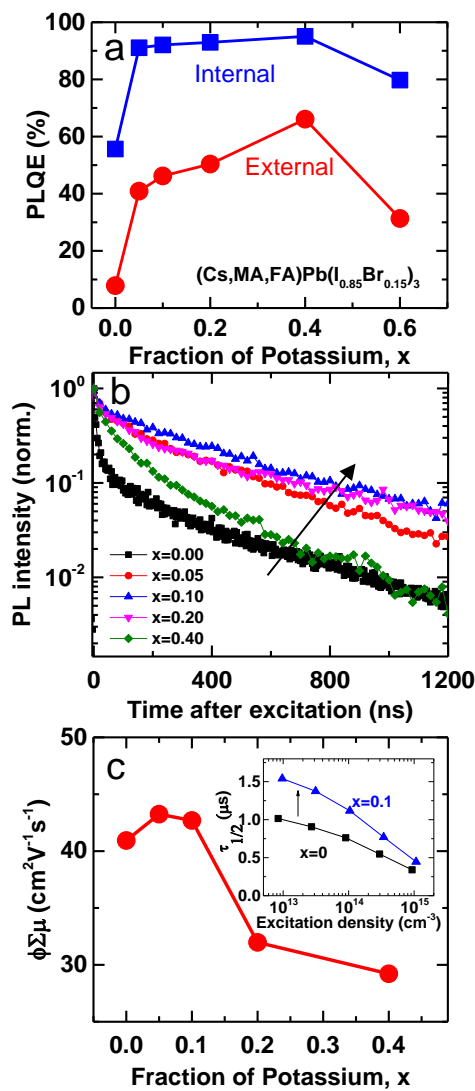

**Figure 1. Increased radiative efficiency and charge carrier mobility through passivation.**
(a) PLQE of passivated perovskite thin films with increasing fraction of potassium measured

under illumination with a 532-nm laser at an excitation intensity equivalent to ~1 sun (~60 mW.cm$^{-2}$) after 300 seconds of illumination. (b) Time-resolved PL decays of the films with excitation at 407 nm and pulse fluence of 0.5 µJ.cm$^{-2}$. (c) Maximum photo-conductance for each of the K contents extracted from TRMC measurements (Extended Data Figure 7). Inset: Half lifetimes of the TRMC decay curves for the $x = 0$ and $x = 0.1$ compositions.

Although other passivation approaches have also been shown to increase the luminescence quantum efficiency of perovskite films, they typically disrupt carrier transport[10,19]. We use time-resolved microwave conductivity (TRMC) to assess the impact of the potassium addition on charge transport in the (Cs,FA,MA)Pb(I$_{0.85}$Br$_{0.15}$)$_3$ perovskite thin films (Extended Data Figure 7)[20]. In Figure 1c, we show the maximum photo-conductance for each of the K contents, providing a measure of the charge mobility for each sample. We find that the carrier mobility remains mostly constant at a large value of ~42 cm$^2$V$^{-1}$s$^{-1}$ for perovskite with $x = 0$ and $x = 0.1$, before dropping for higher K content to ~30 cm$^2$V$^{-1}$s$^{-1}$ ($x = 0.4$). This suggests that, at least up to $x = 0.1$, the charge transport remains unperturbed upon addition of K. From half lifetimes $\tau_{1/2}$ (Figure 1c (inset)) across a range of excitation fluences, we find that the charge-carrier recombination is substantially slower for the $x = 0.1$ composition compared to the $x = 0$ reference, with the low-fluence monomolecular lifetime increases from 1 µs ($x = 0$) to 1.5 µs ($x = 0.1$). These findings are consistent with lower trap densities[13] for the passivated samples. These results collectively suggest that the potassium passivation eliminates almost all non-radiative decay channels while retaining excellent charge transport.

In Figure 2a, we show the PLQE from (Cs,FA,MA)Pb(I$_{0.85}$Br$_{0.15}$)$_3$ thin films as a function of time under continuous laser illumination with intensity equivalent to 1-sun solar illumination. We find a substantial but slow transient rise for the reference film ($x = 0$) associated with photo-induced halide migration[21]. In contrast, the high values of PLQE for the passivated films are stable under continuous illumination, suggesting that the photo-induced migration processes are substantially inhibited. To further investigate the latter claim, we add KI to precursor solutions with higher fractions of Br, which typically show substantial PL shifts due to photo-induced halide segregation and subsequent emission dominated by the low-bandgap iodide-rich components[8]. We show that the PL spectral output of passivated films $(Cs,FA,MA)Pb(I_{0.4}Br_{0.6})_3$ is remarkably stable at the optimal bandgap for perovskite/silicon tandems (1.75 eV)[2] under laser illumination with an intensity equivalent to 1-sun (Figure 2c). In contrast, the same composition without passivation shows substantial red-shifts and bandgap changes over time under the same illumination conditions (Figure 2b). In Figure 2d, we show that this photo-stability is also seen across bromide fractions covering the full range of idealised bandgaps for perovskite-perovskite tandems (1.7-1.9 eV)[2] (see Extended Data Figure 8 for the corresponding spectra without K and other examples). We note that this is the first report showing such exceptional stability in mixed halide compositions across a wide range of bandgaps under both solar illumination and ambient conditions[3].

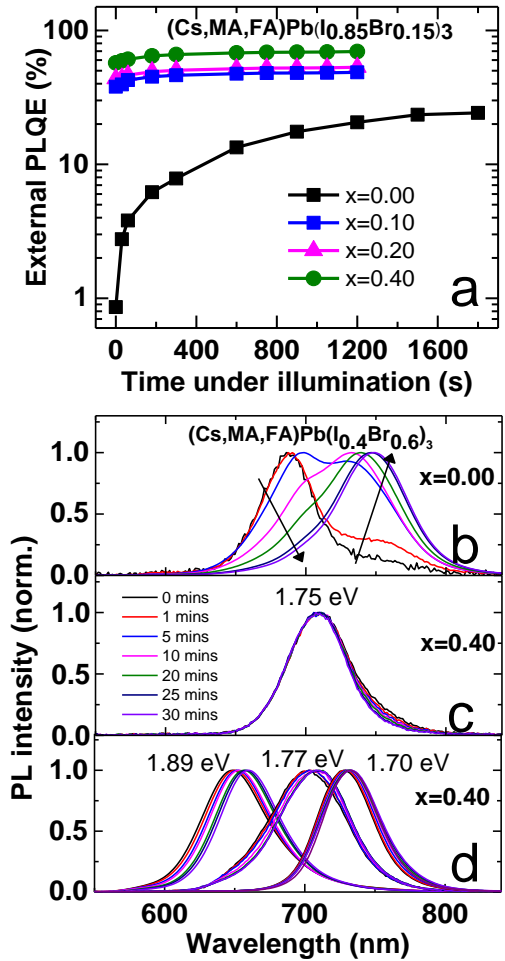

**Fig. 2. Stabilised PLQE and inhibition of photo-induced ion migration.** (a) PLQE for (Cs,FA,MA)Pb(I$_{0.85}$Br$_{0.15}$)$_3$ films illuminated over time with a 532-nm laser at an excitation intensity equivalent to ~1 sun (~60 mW.cm$^{-2}$) in ambient atmosphere. PL from (Cs,FA,MA)Pb(I$_{1-y}$Br$_y$)$_3$ with (b) $y = 0.6$ with passivation ($x = 0$) compared to the (c) unpassivated sample ($x = 0.4$), illuminated continuously in ambient conditions with the same conditions as (a). (d) The PL from the passivated ($x = 0.4$) compositions with $y = 0.4$ (1.70 eV), 0.8 (1.77 eV) and $y = 1$ (1.89 eV), measured over time under the same conditions.

To investigate the local chemical and morphological composition of the (Cs,FA,MA)Pb(I$_{0.85}$Br$_{0.15}$)$_3$ perovskite thin films, we performed scanning transmission electron microscopy-energy dispersive X-ray spectroscopy (STEM-EDX). In Figure 3a, we show a STEM high angle annular dark field (HAADF) cross sectional view of a lamella of the $x = 0.2$ composition. To preserve the perovskite film during specimen preparation, we deposited capping layers of Spiro-OMeTAD and platinum, respectively. From the STEM-EDX elemental analysis, we observe a potassium-rich phase at the grain boundaries of the perovskite as well as the interface with the substrate (Extended Data Figure 9). We also analyzed the STEM-EDX dataset using multivariate analysis methods, specifically a Non-negative Matrix Factorisation (NMF) algorithm[22], which highlights the presence of two compositional phases present in the specimen, reported as Factor 1 and Factor 2 in Figure 3b and c, respectively. Factor 1 shows characteristic EDX features, including Br $L_\alpha$, Pb $M_\alpha$, I $L_\alpha$ lines (Figure 3d), which can be associated with the perovskite phase, while Factor 2 is rich in bromine and potassium (Figure 3e). Interestingly, the signal linked to Factor 2 is particularly strong at the grain boundaries and top and bottom surfaces of the perovskite film. It is likely this is related to a crystalline phase observed in Grazing-Incidence Wide-Angle X-Ray Scattering (GIWAXS) experiments (Extended Data Figure 10). We note that these results are also consistent with Hard X-ray Photoelectron Spectroscopy (HAXPES) measurements, where we observe that the concentration of K decreases when moving from the film surface into the bulk (Extended Data Figure 11). These results collectively indicate the formation of potassium halide (particularly Br-rich) passivation layers decorating the surfaces, with the potassium not incorporating into the perovskite lattice. We note that this is in contrast to other works reporting the addition of small monovalent cation to the perovskite including Na, Rb or K, which propose incorporation of these components into the crystal lattice[5,23–26]. The passivating potassium-halide interlayers we realise in the compositions presented here are uniquely responsible for the exceptional improvement of the material optoelectronic properties.

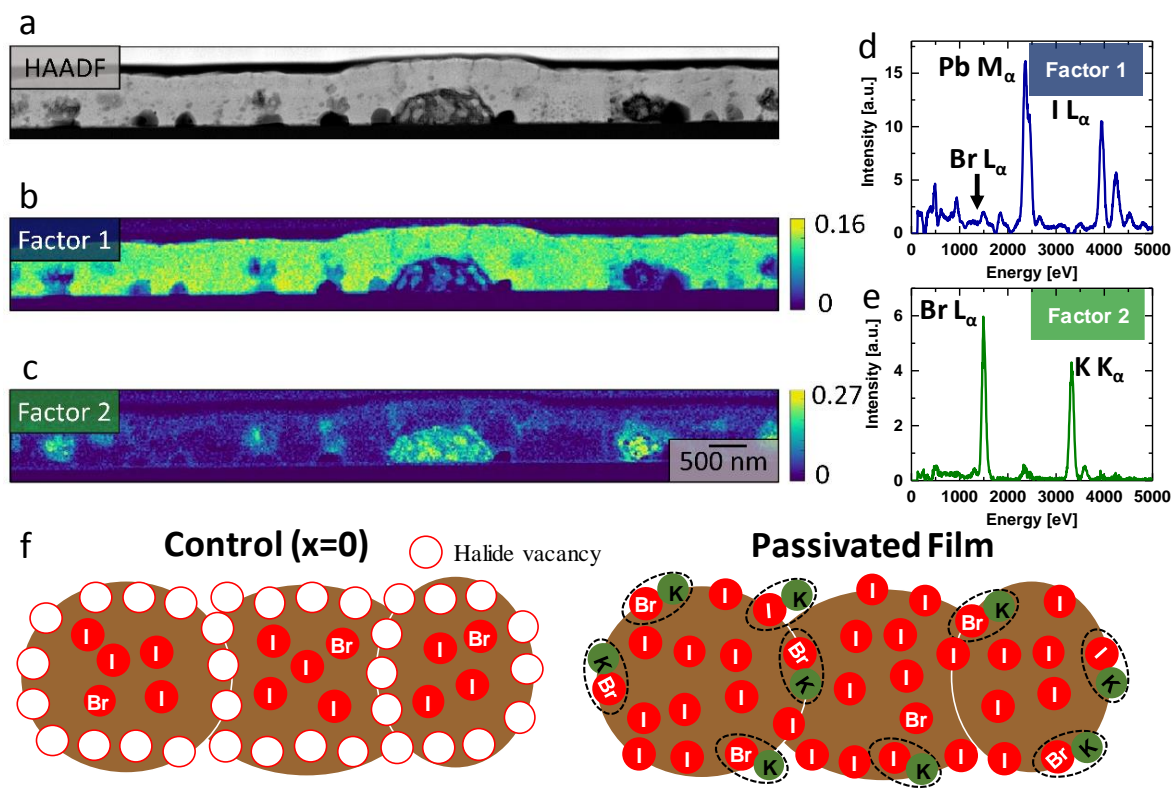

**Figure 3. Cross-section chemical characterization**. (a) HAADF STEM cross sectional image of a $(Cs,FA,MA)Pb(I_{0.85}Br_{0.15})_3$ passivated perovskite thin film ($x = 0.20$). NMF decomposition results in (b) factor 1 associated to the perovskite layer and in (c) factor 2 indicating the presence of a K and Br rich phase. The profiles for (d) factor 1 and (e) factor 2. (f) Schematic of a film cross-section showing halide vacancy management with excess halide, where the surplus halide is immobilised through complexing with potassium into benign compounds at the grain boundaries and surfaces.

In Figure 3f, we summarise our interpretation of the above-mentioned results to propose a mechanism for the enhancements in the optoelectronic properties of perovskite thin films through KI addition. As reported previously, a large density of charge-carrier traps form during the growth of perovskite films, which could originate from halide vacancies particularly at surfaces or grain boundaries[27]. Here, we are introducing excess iodide through the KI source into the perovskite precursor solutions, which compensates for the halide deficiencies in the reference films. The excess halides fill these vacancies, thereby passivating the non-radiative recombination pathways, leading to exceptional internal PLQE values. The excess halides could themselves lead to additional non-radiative decay pathways, for example through iodide interstitials, as well as provide species responsible for photo- and field-induced migration processes. However, the K is able to bind to the excess halide ions, immobilising them as benign potassium-halide species, which are segregated to the grain boundaries and surfaces (cf. STEM-EDX analysis), thus inhibiting halide migration and forming K-halide-rich interlayers. At K content beyond $x = 0.1$, these non-perovskite species are too large and perturb charge transport, leading to decreases in charge carrier mobility as ascertained from TRMC measurements. This suggests there is an optimal K content at $x$ ~0.1, which is a compromise between high radiative efficiency and retention of high charge carrier mobility. Finally, we propose that potassium selectively depletes the Br from the perovskite crystal structure, which is consistent with an increased lattice parameter and red-shifting band-edge with K addition (Extended Data Figure 3). We note that these effects, along with such exceptional optoelectronic properties, are not achieved in the absence of Br (Extended Data Figure 4). The addition of small fractions of bromide to the perovskite precursor solutions has consistently been shown to improve perovskite film formation and resulting properties[28]. However, bromide-rich perovskites typically have increased sub-gap trap states and inferior charge carrier mobility compared to their iodide-based counterparts[29]. Here we demonstrate that we can exploit the beneficial properties of bromide in the grain formation process while suppressing the formation of Br-induced defect states in the bulk of the crystal.

At open-circuit in a solar cell approaching its efficiency limits, external luminescence should be maximised[12], and we must minimise any additional non-radiative losses upon introduction of device electrodes. In Figure 4a, we show the time-resolved PL decays for the (Cs,FA,MA)Pb($I_{0.85}Br_{0.15}$)$_3$ perovskite with and without potassium interlayers when deposited on a standard n-type electron-accepting contact comprised of a thin mesoporous layer of $TiO_2$. We show in Figure 4d the corresponding PLQE measurements with excitation density equivalent to 1 sun. We find that charge-carrier recombination in the presence of the electrode is slowed in the presence of the passivating potassium interlayers, with the PLQE dropping by a factor of only 1.7 (to 27.1%) compared to a drop of a factor of 6.7 (to 3.0%) in the case without the passivating interlayer. This suggests that the potassium interlayer leads to an improved interface with reduced non-radiative recombination. In Figure 4b, we show the time-resolved PL decays for the perovskite containing a top layer of a standard p-type hole-accepting contact Spiro-OMeTAD. We again find that the potassium interlayer leads to slower charge carrier recombination and a less significant drop in PLQE upon introduction of the electrode (Fig. 4d), with a drop in PLQE of factor 4.5 to 10.4% compared to a drop by a factor 38 to 0.5% without the potassium. This indicates that the hole-collecting electrode Spiro-OMeTAD is responsible for the majority of the non-radiative losses in a full solar cell, but these losses can be substantially mitigated in the presence of the potassium-halide interlayers. Finally, we show the time-resolved PL decays for the full device stack (i.e. both electrodes) in Figure 4c, clearly showing the slowed recombination and vastly reduced PLQE drop with the potassium sample. The external PLQE of the full stack is retained at 14.5% with the passivating interlayers (a reduction of factor 3.2 upon addition of the electrodes), an order of magnitude higher than the 1.2% of the control stack (reduction of factor 17). These results show that the potassium interlayers not only improve the optoelectronic properties of the neat material, but also lead to vastly improved interfaces with device electrodes.

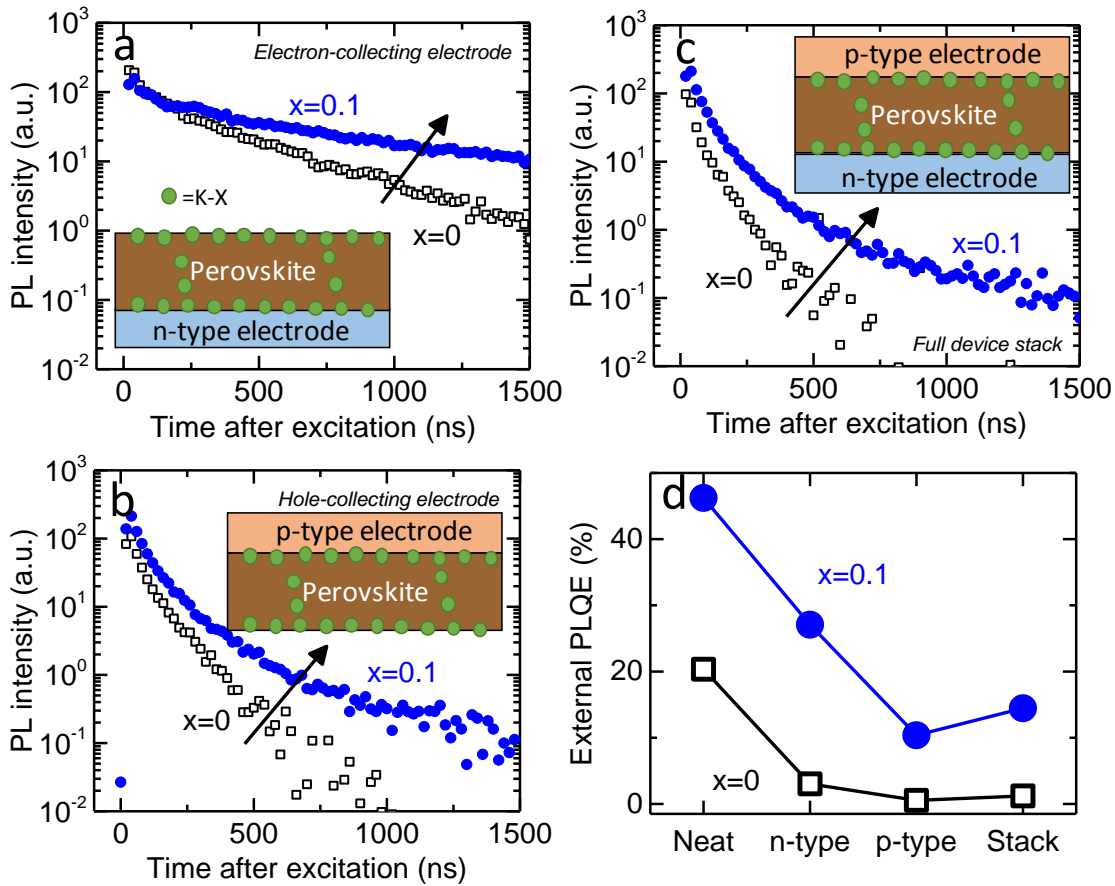

**Figure 4. Luminescence properties of the perovskite when interfaced with solar cell device contacts.** Time-resolved PL decays of encapsulated (Cs,FA,MA)Pb(I$_{0.85}$Br$_{0.15}$)$_3$ films ($x = 0$ and $x = 0.1$) with excitation at 407 nm and pulse fluence of 0.25 μJ.cm$^{-2}$ when the perovskite is interfaced with (a) an n-type electron-collecting electrode (compact-TiO$_2$/thin-mesoporous TiO$_2$), (b) a p-type hole-collecting electrode (Spiro-OMeTAD), and (c) both electrodes in a full device stack. (d) External PLQE measurements of the perovskite in each configuration measured under illumination with a 532-nm laser at an excitation intensity equivalent to ~1 sun (~60 mW.cm$^{-2}$).

To validate our findings in operating solar cells, we construct the full solar cells using the device architecture fluorinated-tin oxide (FTO)/compact-TiO$_2$/thin-mesoporous TiO$_2$/perovskite/Spiro-OMeTAD/Au. In Figure 5a, we show the forward and reverse current-voltage (J-V) curves of champion devices containing the $(Cs,FA,MA)Pb(I_{0.85}Br_{0.15})_3$ absorbers with $x = 0$ and $x = 0.1$ under full simulated sunlight, with the extracted parameters given in Table 1 (see Extended Data Figure 12 and 13 and Table 1 for other K compositions and device statistics). We find that the device efficiency increases from 17.3% ($x = 0$) to 21.5% ($x = 0.1$) with passivation, with the elimination of hysteresis in the latter case. This is consistent with a rapid rise to a stabilized power efficiency of 21.3%, compared to a slower rise to just 17.2% for the control (Fig. 5a inset); these results are also consistent with an inhibition of ion migration related to hysteretic and slow transient effects[30]. We see an increase in open-circuit voltage ($V_{oc}$) with passivation from 1.05 V (x=0) to 1.17 V (x=0.1) (Figure 5c), which is consistent with the superior radiative efficiency (Figure 4d). The total loss-in-potential (difference between $V_{oc}$ and bandgap, 1.56 eV) is only 0.39 V, which surpasses market-leading silicon solar cells (~0.40 V) and matches the lowest loss reported in a perovskite solar cell to date[5]. We also see an increase in the short-circuit current ($J_{SC}$) with K addition up to $x = 0.1$ (Figure 5c), consistent with the increased carrier mobility and lifetime (diffusion length)[31]. The optimal device performance at $x = 0.1$ therefore validates the compromise between radiative efficiency and charge carrier mobility. We conducted preliminary stability tests and found a negligible drop in shelf-life performance of the $x = 0.1$ devices with regular testing over a month, and that the devices retain over 80% of their initial performance after 300 hours of continued operation close to the maximum power point (Extended Data Figure 12). We also show device results for larger bandgap $(Cs,FA,MA)Pb(I_{0.4}Br_{0.6})_3$ absorbers (Fig. 5d and 5f), attaining a PCE of 17.9% with minimal hysteresis for the $x = 0.1$ composition and stabilized power output of 17.1% (see Table 1 for parameters and Extended Data Figure 13 for further device statistics). This is one of the highest efficiencies to date for a large bandgap (1.78 eV) perovskite ideally suited for tandem applications[3].

**Table 1**. Champion photovoltaic device parameters. The bandgaps are extracted from the EQE onsets, and the $V_{oc}$ loss is the difference between the bandgap and $V_{oc}$.

| Potassium fraction, x | Bandgap [eV] | $J_{sc}$ [mA.cm$^{-2}$] | $V_{oc}$ [V] | Fill factor | PCE [%] | $V_{oc}$ loss [V] |
|---|---|---|---|---|---|---|
| (Cs,MA,FA)Pb(I$_{0.85}$Br$_{0.15}$)$_3$ | | | | | | |
| 0.00 | 1.59 | 22.6 | 1.05 | 0.73 | 17.3 | 0.54 |
| 0.10 | 1.56 | 23.2 | 1.17 | 0.79 | **21.5** | **0.39** |
| (Cs,MA,FA)Pb(I$_{0.4}$Br$_{0.6}$)$_3$ | | | | | | |
| 0.00 | 1.83 | 15.3 | 1.12 | 0.72 | 12.3 | 0.71 |
| 0.10 | 1.78 | 17.9 | 1.23 | 0.79 | **17.5** | **0.55** |

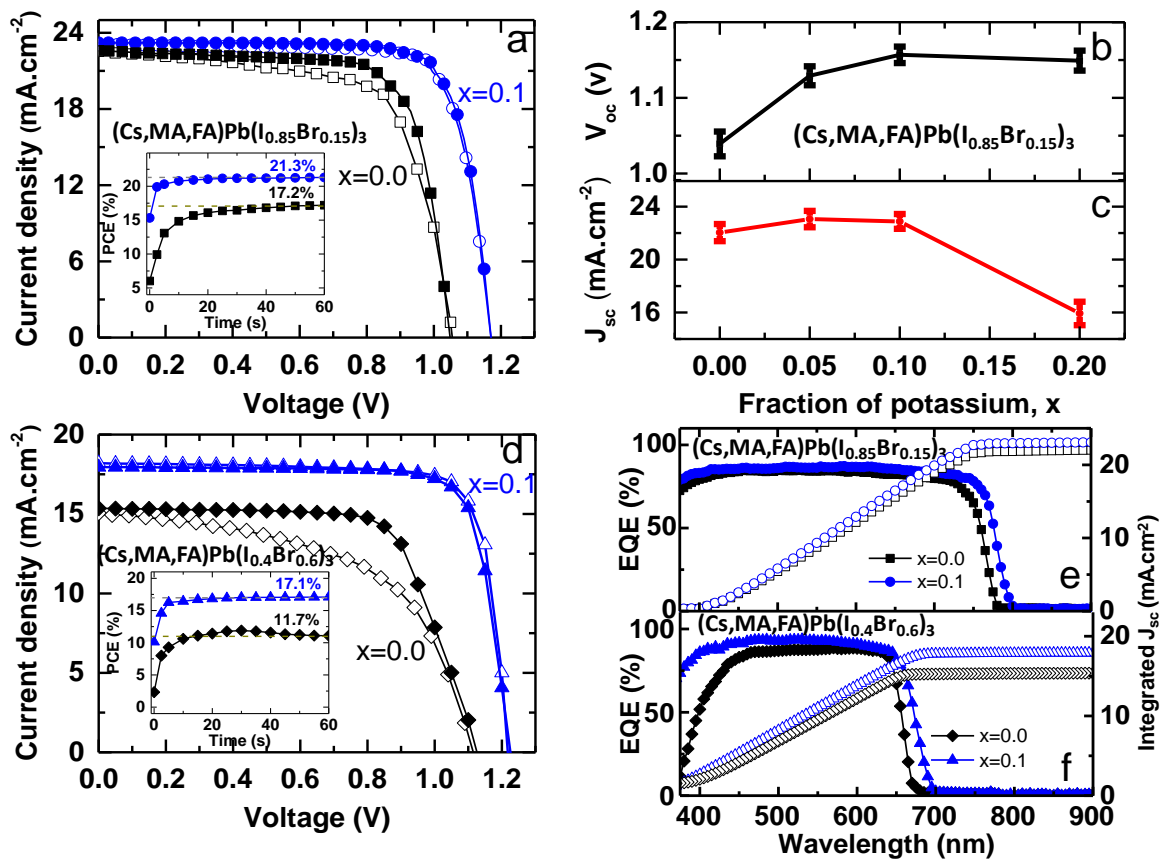

**Figure 5. Enhanced solar cell power conversion efficiency**. (a) Forward (open symbols) and reverse (closed symbols) J-V curves of champion solar cells with (Cs,MA,FA)Pb(I$_{0.85}$Br$_{0.15}$)$_3$ absorbers without (*x* = 0) and with (*x* = 0.1) passivation, measured under full simulated solar illumination conditions (AM1.5, 100 mW.cm$^{-2}$). Inset: Stabilised power output under the same conditions. (b) Open-circuit voltage ($V_{oc}$) and (c) short-circuit current ($J_{SC}$) as functions of *x*, with error bars representing the standard deviation across 10 devices for each composition. (d) J-V curves of champion solar cells with (Cs,MA,FA)Pb(I$_{0.4}$Br$_{0.6}$)$_3$ absorbers without (*x* = 0)

and with ($x = 0.1$) potassium. External quantum efficiencies (EQE) and integrated short-circuit current for the (e) (Cs,MA,FA)Pb(I$_{0.85}$Br$_{0.15}$)$_3$ and (f) (Cs,MA,FA)Pb(I$_{0.4}$Br$_{0.6}$)$_3$ devices.

These results are particularly remarkable for three key reasons. The first is that the perovskite films and interfaces are surprisingly tolerant to passivating additives. Here, we have introduced insulating interlayers at high enough loading to passivate surfaces without compromising charge transport or extraction in the full device. This tolerance is in contrast to conventional semiconductors such as GaAs, which require more complicated approaches to reduce non-radiative surface recombination such as controlled growth on lattice-matched substrates[32]. The second is that these interlayers can stabilise the luminescence and therefore power output in a device across a range of perovskite bandgaps. The third is that these results directly show the importance of obtaining high, stable external luminescence yields in the full device stacks containing luminescent absorber layers capable of recycling photons[12,33]. The internal luminescence yields approaching 100%, along with small loss in external luminescence yield in the full device stack, shows that perovskites can sustain the necessary photon gas to achieve voltage losses low enough to rival GaAs. We expect that further gains in perovskite device performance will be achieved through device optimization, which will include further improvement of the contacts to entirely eliminate the non-radiative losses at the interface. The combination of high radiative efficiency, excellent charge transport and truly photo-stable bandgaps makes these passivation approaches an extremely promising route to take perovskite devices to their efficiency limits across a range of bandgaps.

**Methods**

*Materials.* All the organic cation salts were purchased form Dyesol; the Pb compounds from TCI and CsI and KI from Alfa Aesar. Spiro-OMeTAD was purchased from Borun Chemicals and used as received. Unless otherwise stated, all other materials were purchased from Sigma-Aldrich.

*Film preparation.* The triple-cation-based perovskite $Cs_{0.06}FA_{0.79}MA_{0.15}Pb(I_{0.85}Br_{0.15})_3$ was prepared by dissolving $PbI_2$ (1.2 M), FAI (1.11 M), MABr (0.21 M) and $PbBr_2$ (0.21 M) in the mixture of anhydrous DMF:DMSO (4:1, volume ratios) followed by addition of 5 volume percent from CsI stock solution (1.5 M in DMSO). The potassium iodide was first dissolved in a mixed solution of DMF/DMSO 4:1 (v:v) to make a stock solution of 1.5 M. We then added the KI solution into the triple cation perovskite solution in different volume ratios. We then spin-coated the perovskite solutions using a two-step program at 2000 and 6000 rpm for 10 and 40 seconds, respectively, and dripping 150 µL of chlorobenzene after 30 seconds. We then annealed the films at 100°C for 1 hour. All the film preparations were performed in a nitrogen-filled glove box.

*Device fabrication.* We used the device architecture of fluorinated-tin oxide (FTO)/compact-$TiO_2$/thin-mesoporous $TiO_2$/perovskite/Spiro-OMeTAD/Au. We followed the same procedures for substrate preparation as well as deposition of both electron and hole transport layers (i.e. $TiO_2$, Spiro-OMeTAD) as our previous work[34].

*Optical characterization.* Absorption spectra were recorded with a Perkin-Elmer Lambda 1050 spectrophotometer equipped with an integrating sphere. The thin films were placed in front of the sphere to measure the fraction of transmitted light ($F_T$) and under an angle of 10° inside the sphere to detect the total fraction of reflected and transmitted photons ($F_{R+T}$). From here, we calculated the fraction of absorbed light ($F_A$); $F_A=1-F_{T+R}$.

*Photoluminescence quantum yield.* Perovskite films were placed in an integrating sphere and were photoexcited using a 532 nm continuous-wave laser. The laser and the emission signals were measured and quantified using a calibrated Andor iDus DU490A InGaAs detector for the determination of PL quantum efficiency. The PLQE was calculated as per de Mello et al[35].

*Time-resolved photoluminescence.* Time-resolved photoluminescence measurements were acquired with a gated intensified CCD camera system (Andor iStar DH740 CCI-010) connected to a grating spectrometer (Andor SR303i). Excitation was performed with femtosecond laser pulses which were generated in a homebuilt setup by second harmonic generation (SHG) in a BBO crystal from the fundamental output (pulse energy 1.55 eV, pulse length 80 fs) of a Ti:Sapphire laser system (Spectra Physics Solstice). The pulse energy was varied ranging from 0.05 to 1.5 µJ.cm$^{-2}$. Temporal resolution of the PL emission was obtained by measuring the PL from the sample by stepping the iCCD gate delay relative to the pump pulse. The gate width was 20 ns.

*Time-resolved microwave conductivity (TRMC) measurements.* The TRMC technique evaluates the change in reflected microwave power by the loaded microwave cavity upon pulsed laser excitation. The photo-conductance (ΔG) of the perovskite films was deduced from the collected laser-induced change in normalized microwave power (ΔP/P) by $-K\Delta(t)=\Delta P(t)/P$, where K is the sensitivity factor. The yield of generated free charges $\varphi$ and mobility $\Sigma\mu=(\mu_e+\mu_h)$ were obtained by: $\varphi\Sigma\mu=\Delta G/(I_0\beta eF_A)$, where, $I_0$ is the number of photons per pulse per unit area, $\beta$ is a geometry constant of the microwave cell, $e$ is the elementary charge, and $F_A$ is the fraction of light absorbed by the sample at the excitation wavelength of 500 nm.

*Scanning transmission electron microscopy.* A FEI Helios Nanolab dual beam Focus Ion Beam/ Field Emission Gun - Scanning Electron Microscope (FIB/FEGSEM) was employed to prepare a lamella for STEM imaging and analysis. STEM/EDX data were acquired in FEI

Tecnai Osiris TEM equipped with a high brightness Schottky X-FEG gun and a Super-X EDX system composed by four silicon drift detectors, each approximately 30 mm$^2$ in area and placed symmetrically around the optic axis to achieve a collection solid angle of 0.9 sr. Spectrum images were acquired with a probe current of 0.7 nA, an acceleration voltage of 200 kV, a spatial sampling of 10 nm/pixel and 100 ms/pixel dwell time. Data were acquired with Tecnai Imaging and Analysis (TIA) and analysed with Hyperspy.

*Solar cell characterization.* Current – voltage characteristics were recorded by applying an external potential bias to the cell while recording the generated photocurrent with a digital source meter (Keithley Model 2400). The light source was a 450-W xenon lamp (Oriel) equipped with a Schott-K113 Tempax sunlight filter (Praezisions Glas & OptikGmbH) to match the emission spectrum of the lamp to the AM1.5G standard. Before each measurement, the exact light intensity was determined using a calibrated Si reference diode equipped with an infrared cut-off filter (KG-3, Schott). EQE spectra were recorded as a function of wavelength under a constant white light bias of approximately 5mWcm$^{-2}$ supplied by an array of white light emitting diodes. The excitation beam coming from a 300-W xenon lamp (ILC Technology) was focused through a Gemini-180 double monochromator (Jobin Yvon Ltd) and chopped at approximately 2 Hz. The signal was recorded using a Model SR830 DSP Lock-In Amplifier (Stanford Research Systems). All measurements were conducted using a non-reflective metal aperture of 0.105 cm$^2$ to define the active area of the device and avoid light scattering through the sides.

**Acknowledgments:**

M.A.J. thanks Nava Technology Limited and Nyak Technology Limited for their funding and technical support. S.D.S. has received funding from the European Union's Seventh Framework Programme (FP7/2007-2013) under REA grant agreement number PIOF-GA-2013-622630.

The authors thank the EPSRC for support. The authors thank Diamond Light Source for access to beamline I09 and staff member Tien -Lin Lee as well as Ute B. Cappel (Uppsala University) for assistance during the HAXPES measurements. S.C., C.D and G.D. acknowledge funding from ERC under grant number 25961976 PHOTO EM and financial support from the EU under grant number 77 312483 ESTEEM2. M. A. thanks the president of the UAE's Distinguished Student Scholarship Program (DSS), granted by the Ministry of Presidential Affairs. HR and BP acknowledge support from the Swedish research council (2014-6019) and the Swedish foundation for strategic research.